# A Guide to Global Quantum Key Distribution Networks

Jing Wang and Bernardo A. Huberman
Next-Gen Systems
CableLabs

## Abstract

We describe systems and methods for the deployment of global quantum key distribution (QKD) networks covering transoceanic, long-haul, metro, and access segments of the network. A comparative study of the state-of-the-art QKD technologies is carried out, including both terrestrial QKD via optical fibers and free-space optics, as well as spaceborne solutions via satellites. We compare the pros and cons of various existing QKD technologies, including channel loss, potential interference, distance, connection topology, deployment cost and requirements, as well as application scenarios. Technical selection criteria and deployment requirements are developed for various different QKD solutions in each segment of networks. For example, optical fiber-based QKD is suitable for access networks due to its limited distance and compatibility with point-to-multipoint (P2MP) topology; with the help of trusted relays, it can be extended to long-haul and metro networks. Spaceborne QKD on the other hand, has much smaller channel loss and extended transmission distance, which can be used for transoceanic and long-haul networks exploiting satellite-based trusted relays.

# Introduction

Cryptography has been widely used in modern communication systems to protect three aspects of data security, i.e. confidentiality, integrity, and authentication. Confidentiality prevents the content of a data message from being accessed by unintended recipients. Integrity protects a message from being modified during transmission. And authentication prevents spoofing attacks by verifying the identities of communication parties. All three aspects are protected by data encryption.

Today's cryptographic systems can be divided into two categories, symmetric and asymmetric. Asymmetric cryptography, also known as public cryptography, uses public/private key pairs for encryption, signature, and authentication, whose security is based on the computational complexity of intractable mathematical problems. For example, integer factorization for RSA algorithm, discrete logarithm for Diffie-Hellman key exchange, and elliptic-curve discrete logarithm for elliptic-curve cryptography (ECC). It requires a tremendous amount of computational resources to break a public cryptographic system, which is not feasible for state-of-the-art classical computers. But the same task is not that difficult to perform with quantum computers. It has been proven that today's public key infrastructure (PKI) can be broken in polynomial time by Shor's algorithm using quantum computing. To make things even worse, increasing the key length, which used to be an effective countermeasure to fight against the growing computational power of classical computers, cannot defeat the attacks from a quantum computer, because the required qubit numbers for such an attack scale linearly with the key length.

Symmetric cryptography, on the other hand, uses an identical key for both encryption and decryption. Thanks to its superior security against quantum computing, symmetric cryptography has been widely used in modern communications, with the most prevailing algorithm in use given by the Advanced Encryption Standard (AES). Unlike asymmetric cryptography, however, symmetric cryptography cannot handle the key delivery by itself. Most symmetric cryptographic systems rely on PKI for authentication and key delivery, which makes key distribution the weak link in the security of the whole system. The security of a symmetric cryptographic system relies on the secrecy of its keys, and key distribution thus becomes the target of cyberattacks.

Recently, quantum key distribution (QKD) has become a prominent candidate to counteract the challenge from quantum computing. Different from classical cryptography, where key security is protected by the computational complexity of certain mathematical problems, QKD offers information theoretic security guaranteed by quantum mechanics, i.e., the keys are always safe even if an adversary has unlimited computing power.

It should be emphasized that improvements in data security tend to be incremental. To combat the ever-growing computational power of the attacks, new algorithms are constantly invented to make it harder and harder to break the keys, only to be broken at a future date. QKD, based on the laws of physics is not only a disruptive technology, but one that eliminates the security weaknesses inherent in key delivery.

Figure. 1 shows a global telecommunication network, which can be divided into four segments according to their coverage area. Each segment features different network topologies. For example, intercontinental (>5000 km) and long-haul (1000-5000 km) networks feature point-to-point (P2P) topology; metro (100-1000 km) networks utilize ring and mesh topologies; and access (<100 km) networks have tree or star topologies.

In quantum communications, in order to avoid the photon-number-splitting (PNS) attack, single photon pulses are used to carry qubits. Most photons are lost along the propagation path due to channel loss and only a small portion of them arrive at the receiver. To make things worse, quantum signals cannot be amplified during transmission because of the no-cloning theorem. These factors make quantum communication severely limited in the distance it can cover.

An ideal way to extend the distance of a quantum link is to use quantum repeaters to divide a long path into smaller segments. However, a quantum repeater needs three building blocks, entanglement swapping, entanglement purification, and quantum memories, all of which are far away from practical implementations. While there have been several demonstrations of entanglement swapping over 1000 km, entanglement purification and qubit storage are still immature, with the long storage time and high retrieval efficiency of qubits needed remaining the major challenges.

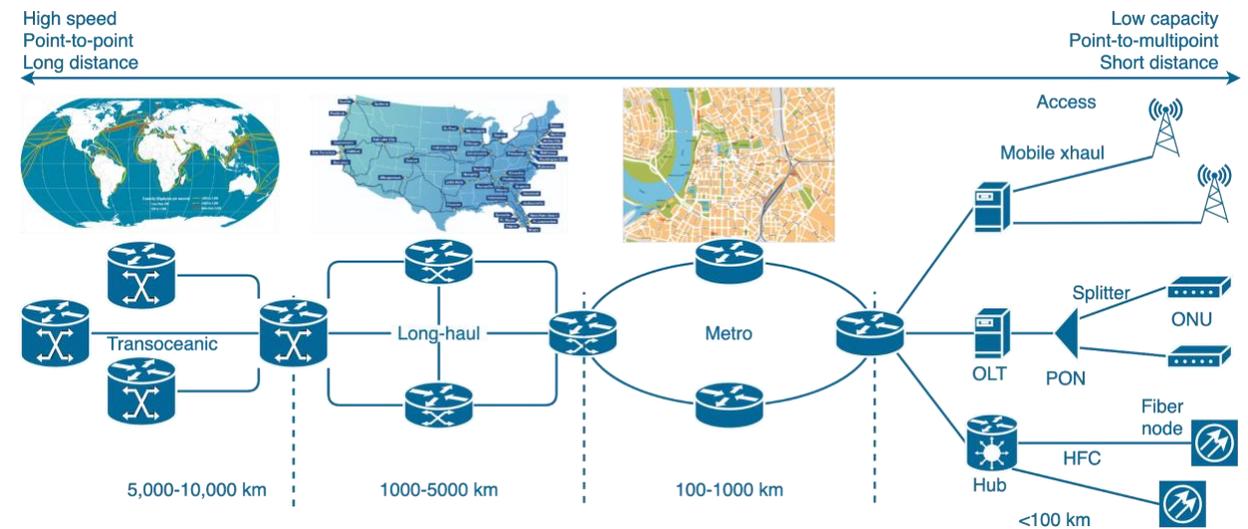

Figure. 1. Global coverage of telecommunication networks, from intercontinental, long-haul, metro, to access networks.

To alleviate the bottleneck created by quantum repeaters, many QKD technologies have been demonstrated, including terrestrial QKD via optical fibers, free-space optics and spaceborne QKD via satellites. However, none of them is able to provide global coverage for quantum key delivery. Terrestrial QKD is limited to short transmission distance. The channel loss of fiber based QKD increases exponentially with the fiber length and the key rate drops dramatically. QKD via free-space optics is subject to environmental influences, such as vibration, adverse weather, atmosphere turbulence, and thus limited to even shorter distances. Spaceborne QKD based on satellites, however, has much lower channel loss and higher key rates and longer transmission distance, with the channel loss limited by diffraction instead of absorption. On the other hand, terrestrial QKD technologies are compatible with point-to-multipoint (P2MP) network topologies; whereas spaceborne solutions surfer from high deployment costs and are only suitable for P2P long-haul applications. To make things worse, spaceborne QKD only works at night with a small coverage area and short time window.

We thus propose to use terrestrial and spaceborne QKD technologies to complement each other in order to enable global coverage of QKD networks from transoceanic to access networks. Since to date there is no investigation that compares the pros and cons of different QKD technologies, we perform a comparative study of various state-of-the-art QKD technologies, including terrestrial technologies via optical fibers and free-space optics, and spaceborne solutions via satellites. Furthermore, we develop the technical selection criteria and deployment requirements that will be needed for different QKD technologies in each segment of the networks.

## Proposed Method and Operation Principles

### Terrestrial QKD via optical fibers

The architecture of fiber based terrestrial QKD networks is shown in Fig. 2. Optical fibers are deployed among network nodes to carry both classical and quantum channels. Due to the weak amplitude of quantum signals, quantum channels are usually deployed in dedicated fiber links to avoid the interference from classical channels. In the case of fiber deficiency, such as in access networks, a quantum channel could also be deployed with classical channels in the same fiber using time/wavelength-division multiplexing (TDM/WDM) techniques, where special care needs to be taken to address the interferences, i.e. Raman scattering noise, from the classical channels. There have been several fiber-based QKD networks demonstrated, e.g., the DARPA network in Boston, the SECOQC network in Vienna, and the Beijing-Shanghai backbone quantum link in China.

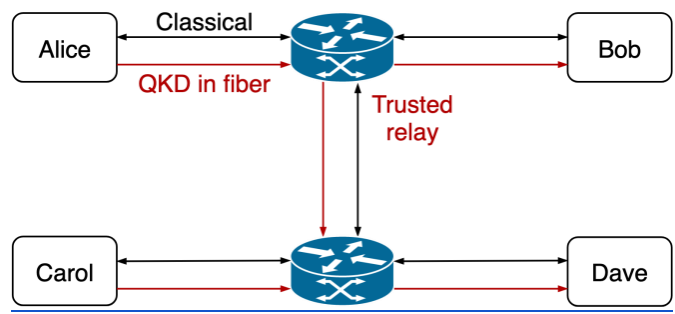

Fig. 2. Terrestrial QKD based on optical fibers

In fiber based QKD networks, quantum channels are confined to the fiber waveguide and shielded from external conditions, such as background light, adverse weather, vibration, and obstructions. Line-of-sight (LoS) connection is not a requirement of fiber-based QKD. On the other hand, absorption in fibers makes the channel loss increase exponentially with distance. Since the key rate is reduce exponentially with channel loss, this makes fiber based QKD technology impractical for long distance applications. For example, with a loss of 0.2 dB/km, a 1000 km fiber introduces a channel loss of 200 dB, which is so high that only 0.3 photons arrive at the receiver per centenary even if a 10-GHz single photon source was used at the transmitter. The current distance record of terrestrial QKD based on optical fibers is 500 km, as reported by Toshiba, whereas the key rate is orders of magnitude lower than practical requirements. In real applications, usable distances will be further reduced to ~100 km.

To extend the reach of fiber based QKD technology, trusted relays are used in order to divide a long path into shorter segments. The Beijing-Shanghai backbone quantum link in China uses 32 trusted relay nodes to divide the overall distance more than 2000 km into many short segments, each less than 100 km. The operation principle of a trusted relay is shown in Fig. 3. The trusted relay node, Charlie, first performs QKD with Alice and Bob, and obtains keys $K_A$ and $K_B$, respectively. It then makes a parity announcement of $K_C = K_A \oplus K_B$, which is a bitwise parity check of the keys $K_A$ and $K_B$. Since the original keys are independent bit strings, their bitwise parity is a uniformly random bit string, which does not reveal any information about the keys. With the help of $K_C$, both Alice and Bob can then infer the key of each other using the fact that $K_A \oplus (K_A \oplus K_B) = K_B$, and $K_B \oplus (K_A \oplus K_B) = K_A$. Notice that since the trusted relay holds all the keys, any access to the relay node gives an adversary complete knowledge of keys. Thanks to the trusted-relay technology, fiber-based QKD offers the compatibility to P2MP network topology, as shown in Fig. 2.

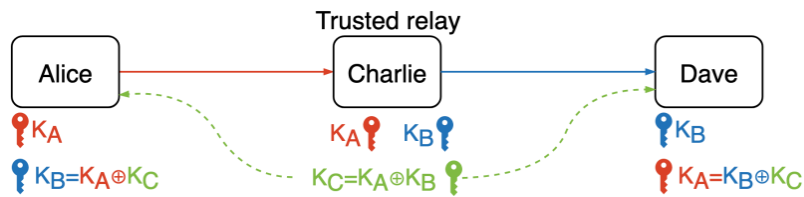

Fig. 3. Trusted relay in terrestrial QKD based on optical fibers

**Terrestrial QKD via free-space optics**

The architecture of terrestrial QKD based on free-space optics is shown in Fig. 4. Different from fiber optics, free-space optics relies on line of sight (LoS) connections, and the transmitters and receivers of quantum channels are usually deployed on top of buildings to avoid obstruction of the optical path. The classical channels exploit the inter-building fibers or rely on free-space optics as well. Since no fiber trenching is required, QKD based on free-space optics features low-cost and easy deployment and enables fast installation during emergencies and disaster recoveries.

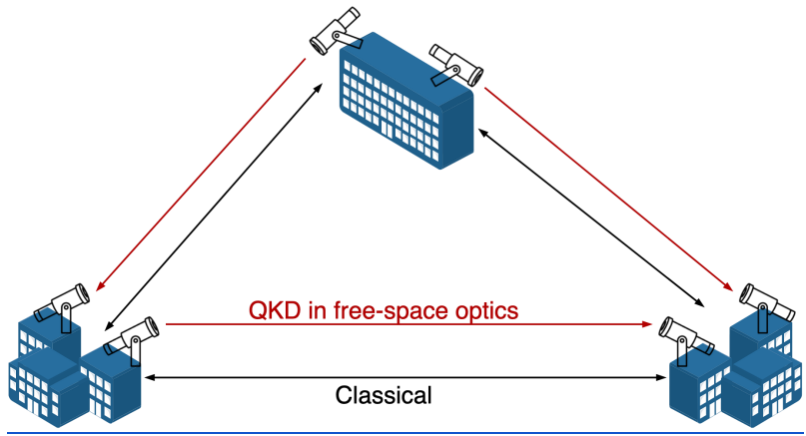

Fig. 4. Terrestrial QKD based on free-space optics

Since the quantum channel is not confined to a fiber waveguide, free-space QKD is sensitive to environmental conditions, such as weather and turbulence. Although the atmosphere has lower absorption than optical fibers, only 0.07 dB/km at 2400 m, the channel loss of a free-space optics link is mainly determined by diffraction, adverse weather (fog, rain, cloud), turbulence, misalignment, and vibration. Quantum channels in free space are also subject to decoherence much more than those in fibers, which further limits the distance to less than 10 km.

On the other hand, free-space optics supports point to multipoint (P2MP) topologies and can easily handle the coexistence of quantum and classical channels. Because the quantum signals propagate in free space, there is no Raman scattering or interference between quantum and classical channels. Limited by short distance, free-space QKD can only be used for inter-building security communication in the last few miles of access networks.

**Spaceborne QKD via satellites as trusted relays**

Fig. 5 shows the architectures of spaceborne QKD via satellites as trusted relays. Satellites used for spaceborne QKD are in the low-earth-orbit (LEO) with altitude less than 900 km, similar to those classical communication satellites in space. For example, Starlink has deployed 844 LEO satellites at the altitude of 550 km and is planning to deploy thousands more in the near future. In Fig. 5(a) classical channels between ground stations are relayed via terrestrial fibers. By exploiting the synergy between the classical and quantum communications in space, the classical channels can also utilize microwave or laser communication among satellites, as shown in Fig. 5(b). Although most Starlink satellites are operating in Ku and Ka bands at present, they are also equipped with laser communication kits for future upgrade.

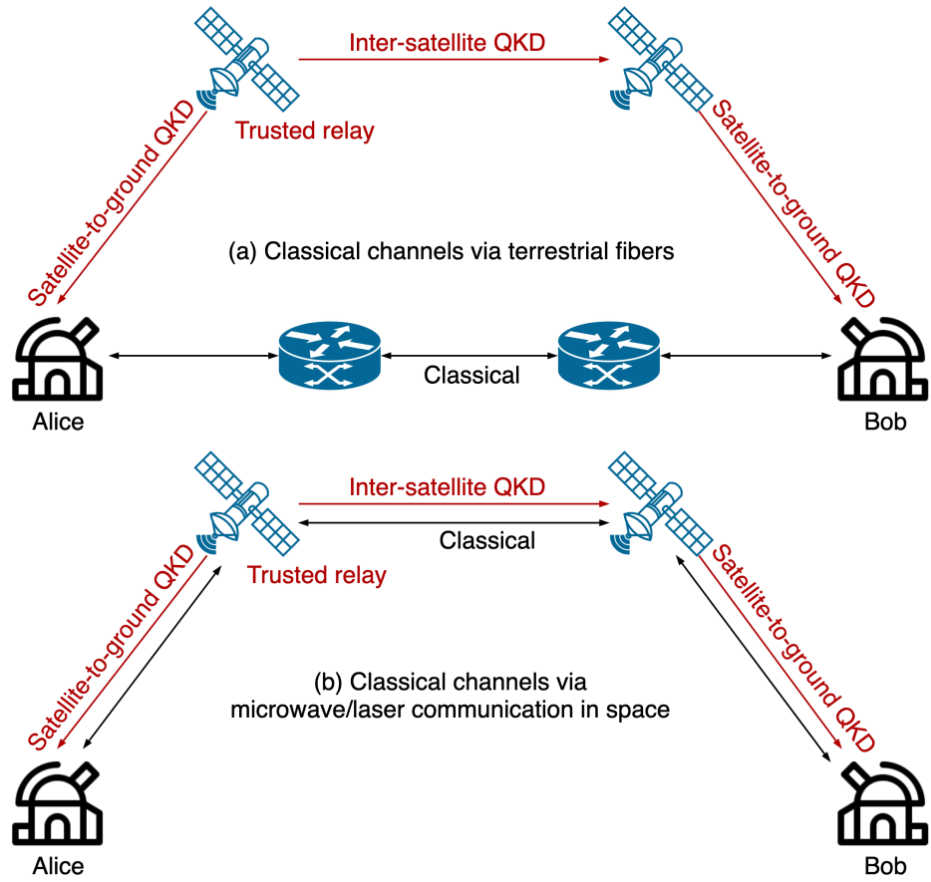

Fig. 5. Spaceborne QKD via satellites. (a) Classical channels via terrestrial optical fibers. (b) Classical channels via microwave or laser communication in space.

Since the effective thickness of the atmosphere is only ~10 km, propagation of a quantum channel takes place mostly in vacuum space, where absorption and turbulence are both negligible. The channel loss of spaceborne QKD is mainly determined by the beam diffraction in the space and scales quadratically with distance. In comparison, the channel loss of terrestrial QKD is caused by absorption and scales exponentially with distance. Therefore, spaceborne QKD has much lower channel loss and extended transmission distance than terrestrial solutions. For example, a 600-km optical fiber introduces 120-dB loss; a free-space link with the same length from satellite to ground only has 50-dB loss if a reasonable aperture size is used at the receiver telescope. Inter-satellite channels have even lower losses due to the absence of atmosphere. Furthermore, channels in space have reduced decoherence than in optical fibers or terrestrial free-space links. Spaceborne QKD is not reliant upon Earth-bound cables and favorable terrain, and therefore can provide coverage for rural underserved areas.

Channel loss in space comes from two sources, beam diffraction, and beam spreading beyond the effects of diffraction. Diffraction loss is a function of the aperture size of the receiver telescope. Beam spreading arises from wavefront aberrations caused by the refractive index inhomogeneities induced by turbulence. There are two categories of turbulence. Small scale turbulence, which induces beam spreading, and large-scale turbulent eddies with size larger than the beam spot, which produce beam wandering.

A long-term beam spot is a superposition of moving short-term beam spots, where the short-term beam size depends on the spreading, and instantaneous beam displacement from the unperturbed position is determined by beam wandering. In real applications, the channel loss of a satellite-to-ground link is dominated by diffraction, followed by beam spreading. Beam wandering has a negligible contribution to losses.

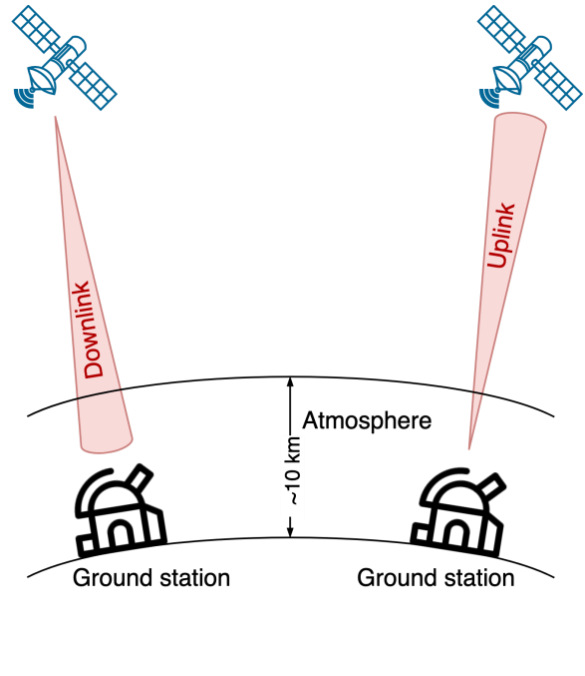

Fig. 6. Downlink (flying transmitter) vs uplink (flying receiver) QKD between a satellite and a ground station.

There are two choices of the direction of the QKD channel between a satellite and ground station, downlink or uplink, i.e., flying transmitter or flying receiver, as shown in Fig. 6. For downlink QKD, the beam first propagates through vacuum space, where the only loss comes from diffraction. The beam then passes through the atmosphere in the final stages of the path, where turbulence-induced wavefront aberrations only spread the beam slightly. Moreover, due to the diffraction effect, the beam arrives at the atmosphere with a size much larger than any turbulent eddy and there is no beam wandering. Overall, atmospheric turbulence has limited impact on the downlink channel loss. For example, the beam size of a downlink channel after 1200 km propagation only expands to 12 m, with diffraction loss of 22 dB depending on the receiver telescope size. The atmospheric turbulence introduces another 3-8 dB attenuation, with an overall channel loss less than 30 dB.

On the other hand, an uplink channel has the beam first propagate through the atmosphere, where the initial wavefront aberration induced by turbulence causes significant beam spreading. The beam size of an uplink channel can reach up to 50 m at 500 km altitude, much larger than the available telescope aperture on a satellite. Downlink channels can exploit a receiver telescope with large aperture size to reduce diffraction loss; whereas uplink channels have a limitation on the telescope size, since large telescopes are too heavy and expensive to be deployed on satellites. Limited by the strong wavefront aberration, large beam spot, and small telescope, uplink channel loss from ground to a satellite at 500 km altitude is more than 50 dB; whereas a downlink channel with the same length only has loss less than 20 dB. Actually, most uplink channels cannot work without the help of the decoy-state technique.

For a QKD link, single-photon detectors at the quantum receiver are the most expensive and delicate devices; whereas the quantum transmitter only requires commercially available off-the-shelf devices. Uplink QKD utilizes a flying receiver, where the sensitive SPDs have to go through adverse conditions in space, including launch vibration, shock in the flight, and extreme temperature variations. Due to sunlight, the temperature on a satellite can vary by up to tens of degrees in a single orbit. Since there is limited electrical power on the satellite for cooling purposes, the only effective way to dissipate the heat is via radiation. To make things worse, most SPDs are avalanche photon detectors (APD), which are vulnerable to dark counts caused by ionizing radiation in space. This is another reason why downlink channels are preferred, so that satellites equipped with a low-cost commercial transmitter are launched into space, whereas expensive and delicate SPDs are properly protected and cooled at ground stations.

Since no amplification, reception, or retransmission is allowed in a quantum channel, there is no way to increase the signal power. The only way to achieve high signal-to-noise ratio (SNR) is to reduce the channel loss and background noise. Downlink channels have much larger SNR than uplink ones thanks to the low loss. In daytime, the background noise is dominated by sunlight, which makes SNR less than 1 and prohibits spaceborne QKD. One possible way to improve daytime SNR is to use wavelengths at Fraunhofer lines, i.e., the absorption lines of the Sun. At night, background noise is contributed by moonlight and scattered light from human activities, which depends on the location of ground stations. SNR at night is orders of magnitude higher than that in the daytime, so downlink at night is the best choice for spaceborne QKD. There are several techniques to improve SNR, e.g., reducing beam size, reducing field-of-view of the receiver telescope, utilizing narrow band filtering and short gating window before the single photon detectors.

Fig. 7 shows the operation principles of spaceborne QKD with satellites as trusted relays. A LEO satellite performs downlink QKD with two ground stations, Alice and Bob, respectively. It then makes a parity announcement, so that both Alice and Bob can infer each other's key. The relay satellite needs LoS connections with both Alice and Bob, but not necessarily at the same time. It can exchange keys with several ground stations one after another as it flies over. Since the relay satellite holds all the keys, any access to the satellite leaks the complete infmroation of keys.

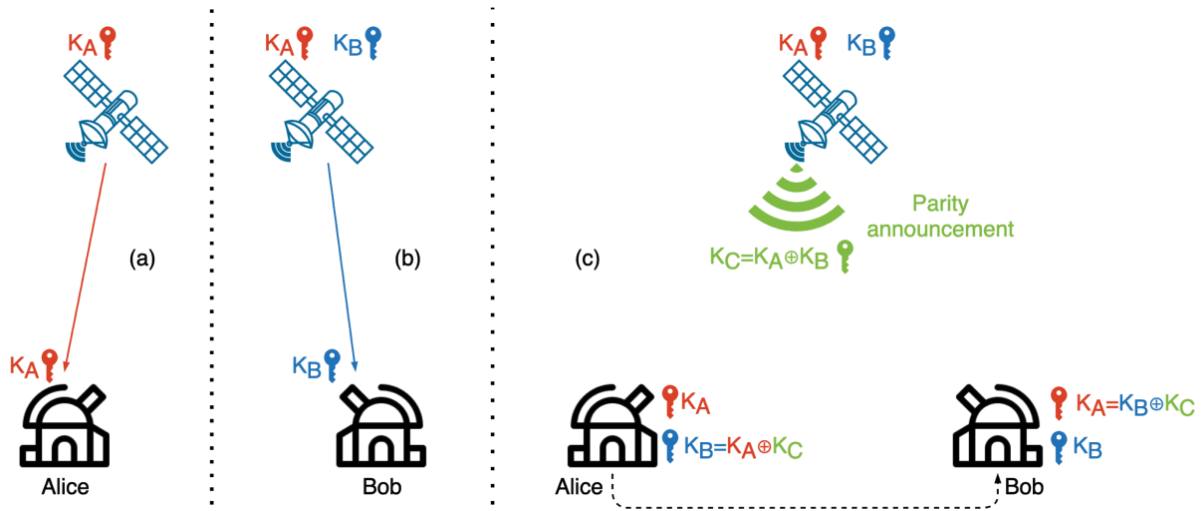

Fig. 7. Spaceborne QKD via satellites as trusted relays. The relay satellite first exchanges keys with ground station Alice (a) and Bob (b), respectively. (c) The satellite makes a parity announcement of two keys, so that both Alice and Bob can infer each other's key.

**Spaceborne QKD via satellites as untrusted relays**

To avoid the potential key leakage in the satellites, untrusted relaying is preferred since the eavesdropper gets no information even if she takes full control of the satellite. Fig. 8 shows two architectures of spaceborne QKD via satellites as untrusted relays, corresponding to downlink and uplink quantum channels, respectively. In Fig. 8(a), an entangled photon source on a satellite is shared by two ground stations. The entangled photons are sent downward from the satellite to Alice and Bob, respectively, who perform random and independent measurements on the incoming photons and use BBM92 or E91 protocol to extract an identical copy of the key. Since the photon source has no control over the qubits carried on entangled photons, the satellite has no information about the final key.

An uplink version of untrusted relay is shown in Fig. 8(b). For measurement-device- independent (MDI)-QKD, two ground stations prepare random qubits independently and send them upwards to the satellite for a Bell-state measurement (BSM). The satellite announces the results of the successful BSM and then Alice and Bob establish an identical copy of key from these events. Further classical communication between Alice and Bob is required than entangled photon pair protocols. The post-selection of successful BSM events actually entangles the qubits from Alice and Bob, making MDI-QKD equivalent to a time-reversed entangled photon pair protocol. MDI-QKD removes all loopholes at the detection side, so the eavesdropper has no information about the key even if she takes control of the satellite.

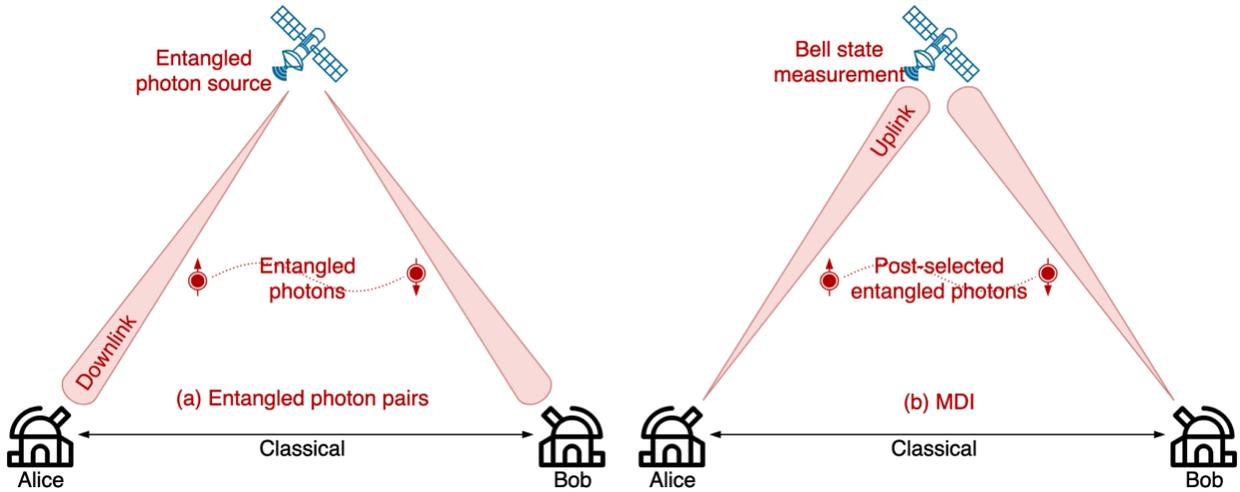

Fig. 8. Spaceborne QKD via satellites as untrusted relays. (a) A satellite distributes entangled photons to two ground stations. (b) Measurement-device independent (MDI) QKD.

For the downlink distribution of entangled photons, channel losses of both arms have to be combined since only photon pairs that arrive at both ground stations can be used. Similarly, for uplink MDI-QKD, only photons from two ground stations that both arrive at BSM are taken into account. Due to the high loss of uplink channels, MDI-QKD via satellite is still not practical. So far, entanglement-based spaceborne QKD is the only way to establish secure keys between two distant locations without a trusted relay. Unlike trusted relaying, untrusted relaying requires *simultaneous* LoS connections from both ground stations to the satellite, which places a limitation on the separation distance between two stations. The current record is 1200 km, achieved by the Micius satellite placed in orbit by China. Wider separation makes a lower slant angle and longer propagation in the atmosphere, which in turn leads to higher channel loss.

**Global coverage of QKD networks**

Table 1 lists the comparison between various QKD technologies, including terrestrial and spaceborne. Terrestrial QKD via optical fibers suffers from high channel loss and short transmission distance but offers the compatibility to a P2MP network topology. Since the quantum channels are confined to the fiber waveguide, QKD via fiber can operate under adverse environmental conditions, such as background light, weather, and vibration. Without relays, it can only be used for access networks with distance less than 100 km.

Trusted relays could be used to extend the reach of fiber based QKD by increasing the number of hops. An interesting synergy with classical fiber communication is that today's fiber cables have repeaters every 100 km, where reamplification, reshaping, and retiming of classical optical pulses are performed. This span is consistent with the distance limit of fiber based QKD, so trusted relay nodes could be deployed at the locations of repeaters to enable coverage for metro and long-haul networks. Since all fiber repeaters have fixed locations, the relay nodes deployed with repeaters are subject to constant surveillance and probing. Spaceborne QKD using satellites as trusted relays provide a more secure medium because the satellite and quantum links are moving fast, making side-channel attacks difficult. Although the atmosphere has lower absorption, the channel loss of free-space QKD is dominated by diffraction, adverse weather and turbulence. The high channel loss limits the distance of free-space QKD to less than 10 km. On the other hand, it supports P2MP topology and can easily handle the coexistence of quantum and classical channels without interference, which makes it suitable for the last few miles among buildings in access networks.

Table 1. Comparison of different QKD technologies.

| QKD technologies | Terrestrial | | Spaceborne | |
| --- | --- | --- | --- | --- |
| | Optical fiber | Free-space optics | Satellite as trusted relay | Satellite as untrusted relay |
| Source of attenuation | Absorption | Diffraction Absorption Weather and turbulence | Diffraction | |
| Other effects | Raman scattering noise from classical channels Immune from environments | Misalignment vibration | Weather Turbulence | |
| Channel loss | High Scale exponentially with fiber length | High Scale exponentially with distance | Low Scale quadratically with distance | |
| Distance | 100 km Unlimited with trusted relay | <10 km | Unlimited | <1000 k |
| P2MP | P2MP | P2MP | P2P | P2P |
| LoS | No | Yes | Yes | Simultaneous LoS with both ground stations |
| Time window | Whole day | Night | Short time window at night (burst) | |
| Deployment cost | Low cost Dedicated fiber or reuse existing ones | No fiber trenching Low cost, simple and fast | Synergy with spaceborne laser communication Expensive and slow | |
| Application scenarios | Long-haul, metro, access, last few miles | Last few miles | Transoceanic, long-haul, metro | Metro |

Compared with terrestrial QKD technologies, spaceborne QKD has lower channel loss and extended transmission distance. Downlink channels from satellite to ground are preferred due to the lower loss and less wavefront abbreviation induced by turbulence. From the perspective of cost and complexity, downlink channels are also preferred since the satellites only need to be equipped with quantum transmitters consisting of commercial off-the-shelf devices, and the expensive and delicate quantum receivers are left on ground.

Spaceborne QKD requires LoS connections between the satellite and ground stations, and only works at night due to the high background noise from the sun during daytime. To reduce the channel loss, LEO satellites are preferred, but the low altitude makes the coverage area small and the fast movement of satellites limits the flyover time for each ground station. Satellites at higher orbit provide wide coverage area and long flyover time, but with the penalty of high channel loss and low key rate. To choose an appropriate altitude, a trade-off has to be made between the coverage area and time window versus channel loss and achievable key rate. An extreme example is a geostationary orbit (GEO) satellite with a fixed location above the equator. It has an operational window of the whole night, but the high altitude of 35,786 km makes the channel loss too high to implement QKD.

Classical communication in space also exploits LEO satellites at the altitude of 300-1000 km. The company *Starlink* plans to deploy thousands of satellites at an altitude of 350-580 km. Although these satellites still use microwave communication in the Ku and Ka bands, most of them are equipped with optical transceivers for future space laser communications. To take advantage of the synergy between quantum and classical communication, classical communication satellites can also be used as QKD trusted relays by adding quantum transmitters onboard. Since quantum transmitters only need commercial off-the-shelf devices, this upgrade will not significantly increase the cost of the satellites. Moreover, the expensive beam acquisition, tracking, and pointing systems designed for classical laser communication can be reused by quantum channels. Spaceborne QKD via satellites as trusted relays has no distance limit and can thus provide coverage for transoceanic, long-haul, and metro networks.

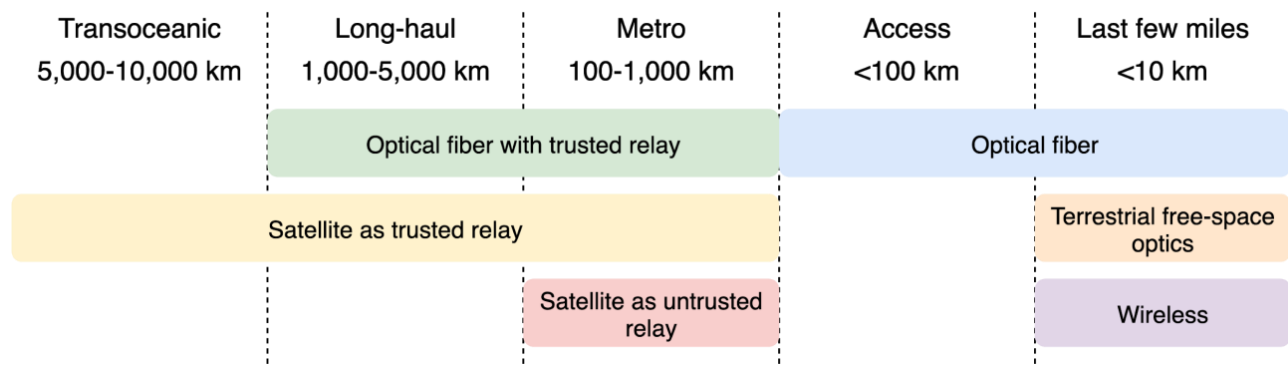

Fig. 9. Deployment strategies for global coverage of QKD networks.

Spaceborne QKD via satellites as untrusted relays shares similar pros and cons with the trusted relays but offers better security since there is no key information leakage at the satellite level. As a penalty, it requires simultaneous LoS connections from the satellite to both ground stations, which limits the separation between two ground stations to less than 1000 km. Therefore, QKD via satellite as untrusted relay can only be used for metro networks. Fig. 9 shows the deployment strategies for global coverage of QKD networks, from transoceanic, long-haul, metro to access networks.

It should be noted that not all user devices are connected with optical fibers or free-space optics. Radio access is used extensively in the last few miles of access networks. In these cases, keys have to be distributed wirelessly in a classical way to user devices. Fig. 10 shows a hierarchical key delivery architecture. Secure sites, e.g., bank buildings, business campuses, government offices, are connected with optical fibers or free-space optics, in which terrestrial QKD links are deployed to deliver keys in an absolutely secure way. Within each secure site, however, the keys are distributed to mobile users wirelessly. This is a trade-off between security and mobility, because it is not feasible to connect all devices with optical fibers, we have to leverage the ubiquity and flexibility of wireless communication. Once the mobile users get the keys, they can use these keys to encrypt their wireless communication. They can even roam away from the secure site and continue their secure communication as soon as they obtain their keys. Once they consume all the keys, they have to return to a secure site to fetch new keys. In this hierarchical architecture, two different levels of security-as-a-service are provided, i.e., absolute security over long distance among secure sites, and computationally classical security over short distance within each site.

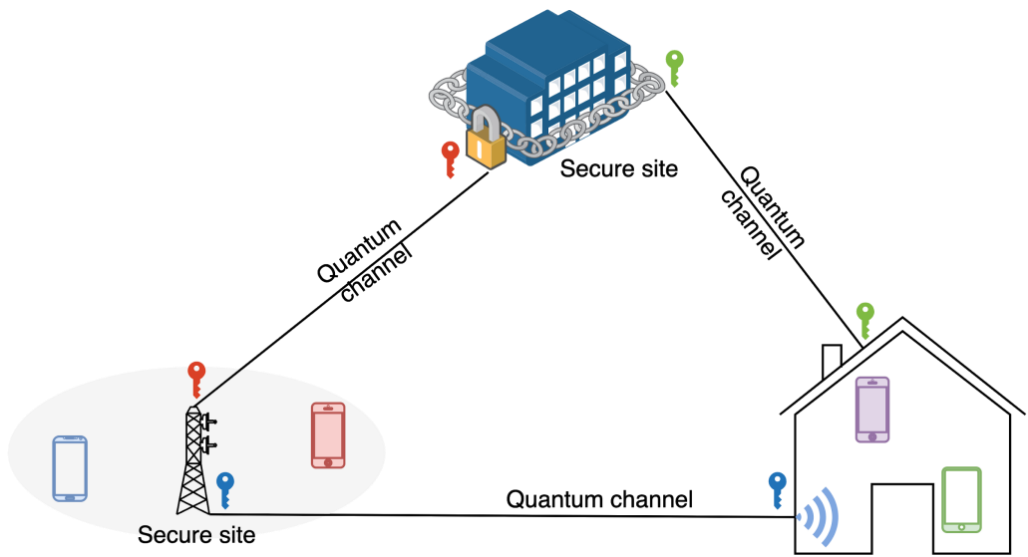

Fig. 10. Hierarchical key delivery in the last few miles.

## Conclusions

To date, although many QKD technologies have been demonstrated, no single one has the capability to enable global coverage of quantum networks. Moreover, there has been no comparative study on the pros and cons of the state-of-the-art QKD technologies that could enable such a deployment. In this paper, a systematic comparison among various QKD solutions is carried out for the first time, in terms of channel loss, potential interference, distance, connection topology, cost and application scenarios. Instead of competing with each other, we first propose to use terrestrial and spaceborne QKD technologies to complement each other in order to enable the global coverage of QKD networks. The systems and methods of using different QKD solutions in various network segments, from transoceanic, long-haul, to metro and access networks, were analyzed, and the selection criteria and deployment requirements for each network segment developed.

Given its compatibility with P2MP topology and 100-km distance limit, terrestrial QKD via optical fibers is suitable for access networks. With the help of trusted relays and by exploiting the synergy with classical optical communication, it can be extended to metro and long-haul networks, where the relay nodes can be deployed at the locations of classical repeaters. Terrestrial QKD via free-space optics is limited to 10-km due to diffraction, weather and turbulence, and is usable in the last few miles of access networks. Spaceborne QKD features low channel loss, reduced decoherence, and extended reach upto ~1000 km. By utilizing satellites as trusted relays, its distance can be extended infinitely. Moreover, spaceborne QKD is not restricted by Earth-bound cables or favorable terrain and provides better coverage for rural underserved areas. We propose to use spaceborne QKD via satellites as trusted relays for transoceanic, long-haul, and metro networks. Finally, QKD solutions via satellites as untrusted relays require simultaneous LoS connections from the satellite to both ground stations and are suitable for metro networks.

## References

1. H. J. Kimble, "The quantum Internet," Nature, vol. 453, pp. 1023–1030, Jun. 2008.

2. S. Wehner, D. Elkouss, and R. Hanson, "Quantum internet: A vision for the road ahead," Science, vol. 362, no. 6412, Oct. 2018.

3. D. Bacco, I. Vagniluca, B. Da Lio, N. Biagi, A. Della Frera, D. Calonico, C. Toninelli, F. S. Cataliotti, M. Bellini, L. K. Oxenløwe, A. Zavatta, "Field trial of a finite-key quantum key distribution system in the Florence metropolitan area" arXiv:1903.12501

4. G. Vallone, D. Bacco, D. Dequal, S. Gaiarin, V. Luceri, G. Bianco, and P. Villoresi, "Experimental Satellite Quantum Communications", https://arxiv.org/pdf/1406.4051.pdf

5. S. Liao, W. Cai, W. Liu, et al., "Satellite-to-ground quantum key distribution," Nature, vol. 549, pp. 43-47, 2017.

6. S. Liao, W. Cai, J. Handsteiner, et al., "Satellite-Relayed Intercontinental Quantum Network," Physical Review Letters, vol. 120, no. 3, pp. 030501, 2018.

7. I. Khan, B. Heim, A. Neuzner and C. Marquardt, "Satellite-Based QKD," Optics and Photonics News, Feb 2018.

8. R. Bedington, J. M. Arrazola, A. Ling, "Progress in satellite quantum key distribution," Nature Partner Journals (NPJ) Quantum Information, vol. 3, no. 30, 2017.

9. H. Takenaka, A. Carrasco-Casado, M. Fujiwara, et al., "Satellite-to-ground quantum-limited communication using a 50-kg-class microsatellite," Nature Photonics, vol. 11, pp. 502-508, 2017.

10. C. Bonato, A. Tomaello, V. D. Deppo, et al., "Feasibility of satellite quantum key distribution," New Journal of Physics, vol. 11, no. 4, pp. 045017, 2009.

11. T. Scheidl, J. Handsteiner, D. Rauch, R. Ursin, "Space-to-ground quantum key distribution," Proceedings vol. 11180, International Conference on Space Optics (ICSO) 2018, Chania, Greece.

12. A. Villar, A. Lohrmann, X. Bai, et al., "Entanglement demonstration on board a nano-satellite," Optica, vol. 7, no. 7, pp. 734-737, 2020.

13. J. Yin, Y. Cao, Y.-H. Li, et al., "Satellite-based entanglement distribution over 1200 kilometers," Science, vol. 356, no. 6343, pp. 1140-1144, 2017.

14. L. Jiang, J. M. Taylor, K. Nemoto, W. J. Munro, R. Van Meter, and M. D. Lukin, "Quantum repeater with encoding," Phys. Rev. A, vol. 79, p. 032325, 2009. [Online]. Available: http://link.aps.org/doi/10.1103/PhysRevA.79.032325

15. S. Muralidharan, L. Li, J. Kim, N. Lütkenhaus, M. D. Lukin, and L. Jiang, "Optimal architectures for long distance quantum communication," Scientific Reports, vol. 6, p. 20463, 2016.

16. C. Elliott, "Building the quantum network," New Journal of Physics, vol. 4, p. 46, 2002.
17. M. Caleffi, A. S. Cacciapuoti, and G. Bianchi, "Quantum internet: from communication to distributed computing!" in Proc. of IEEE/ACM NANOCOM, 2018
18. R.V.MeterandJ.Touch,"Designing quantum repeater networks,"IEEE Communications Magazine, vol. 51, no. 8, pp. 64–71, Aug. 2013
19. A. S. Cacciapuoti et al. The Quantum Internet: Networking Challenges in Distributed Quantum Computing, https://arxiv.org/pdf/1810.08421.pdf

**Acronyms**

| | |
|---|---|
| AES | advanced encryption standard |
| APD | avalanche photon detector |
| BSM | bell state measurement |
| ECC | elliptic-curve cryptography |
| GEO | geostationary orbit |
| LEO | low earth orbit |
| LoS | line-of-sight |
| MDI | measure-device-independent |
| P2MP | point-to-multipoint |
| P2P | point-to-point |
| PKI | public key infrastructure |
| PNS | photon-number-splitting |
| QKD | quantum key distribution |
| SNR | signal-to-noise ratio |
| TDM | time-division multiplexing |
| WDM | wavelength-division multiplexing |